%% file: ICRC2025_PBRoverview.tex
\documentclass[a4paper,11pt]{article}
\usepackage{pos}
\usepackage{units}
\usepackage{subfig}
\title{POEMMA-Balloon with Radio: An Overview}

\author*[a]{J.Eser}
\author[a]{A.V.Olinto}
\author[b]{G.Osteria}

\affiliation[a]{Columbia Astrophysics Laboratory, Columbia University, New York, NY, USA}
\affiliation[b]{Istituto Nazionale di Fisica Nucleare - Sezione di Napoli, Italy}

\onbehalf{for the JEM-EUSO collaboration} 
\emailAdd{jbe2130@columbia.edu}

\abstract{
The POEMMA-Balloon with Radio (PBR) is an Ultra Long Duration Balloon payload scheduled for launch in Spring 2027 from Wanaka, New Zealand. It will circle over the Southern Ocean for a mission duration as long as 50 days, serving as a precursor to the dual satellite mission, Probe of Extreme Multi-Messenger Astrophysics (POEMMA). The PBR mission represents a significant step towards establishing a space-based multi-messenger observatory.

Observations from space will enhance the statistics of the highest-energy cosmic rays and complement ground-based observatories by enabling simultaneous observations of both hemispheres with a single instrument. Additionally, POEMMA will facilitate the measurement of Very-High-Energy Neutrinos (VHENs) following multi-messenger alerts of astrophysical transient events, such as gamma-ray bursts.
PBR is an adaptation of the POEMMA mission, featuring three primary science goals:
\begin{enumerate}
    \item Observe Ultra-High-Energy Cosmic Rays (UHECRs) via the fluorescence technique from suborbital space.
    \item Observe horizontal high-altitude air showers (HAHAs) with energies exceeding the cosmic ray knee (E > 3 PeV) using optical and radio detection for the first time.
    \item Follow astrophysical event alerts in the search for VHENs.
\end{enumerate}
This contribution provides an overview of the PBR payload and discusses the expected performance of its various detectors.
}

\ConferenceLogo{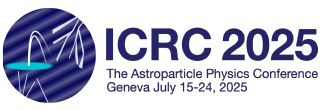}

\FullConference{39th International Cosmic Ray Conference (ICRC2025)\\
 15–24 July 2025\\
Geneva, Switzerland\\}

\begin{document}
\maketitle

\input{Intro}
\input{PBR-Payload}
\input{Summary}

\bibliography{icrc2025_pbr}

%
\input{JEM-EUSO_Authors_August2025}

\end{document}

%% file: Intro.tex
\section{Introduction}
\label{sec:intro}

Over the last decades, significant progress has been made in the study of Ultra-High-Energy Cosmic Rays (UHECRs, $E > 1~\mathrm{EeV}$) and Very High Energy Neutrinos (VHENs, $E > 1~\mathrm{PeV}$). The spectra and compositions of UHECRs have been measured with high precision by the largest ground-based experiments to date: the Pierre Auger Observatory in the Southern Hemisphere and the Telescope Array in the Northern Hemisphere. Both experiments have also found initial hints of anisotropy in the arrival directions of these particles. Yet, despite these advances, several questions remain unanswered, including the cosmological evolution of sources as well as the production and acceleration mechanisms of these particles.

In 2022, the community convened as part of the Snowmass process to propose a roadmap for addressing these open questions over the next two decades. A detailed discussion of the current status and future of UHECR research, including the role of multi-messenger astrophysics, is provided in \cite{Coleman:2022abf}. The community concluded that two complementary types of next-generation detectors are needed to collect the required data: (1) high-accuracy detectors and (2) detectors that maximize exposure at the highest energies.

An obvious choice for maximizing exposure and achieving full-sky coverage is a space-based detector, such as the proposed dual-satellite mission Probe for Multi-Messenger Astrophysics (POEMMA) \cite{POEMMA:2020ykm} or, more recently, M-EUSO \cite{Zbigniew:ICRC}. Such missions would observe neutrinos by detecting the Cherenkov light emitted by extensive air showers (EASs). These upward-moving EASs are produced when a charged lepton, generated after a neutrino traverses the Earth and interacts near the surface, decays in the atmosphere. UHECRs are detected via the fluorescence light emitted by EASs initiated by the interaction of primary cosmic rays with the atmosphere.

\begin{figure}[h!]
   \centering
    \includegraphics[width=.75\textwidth]{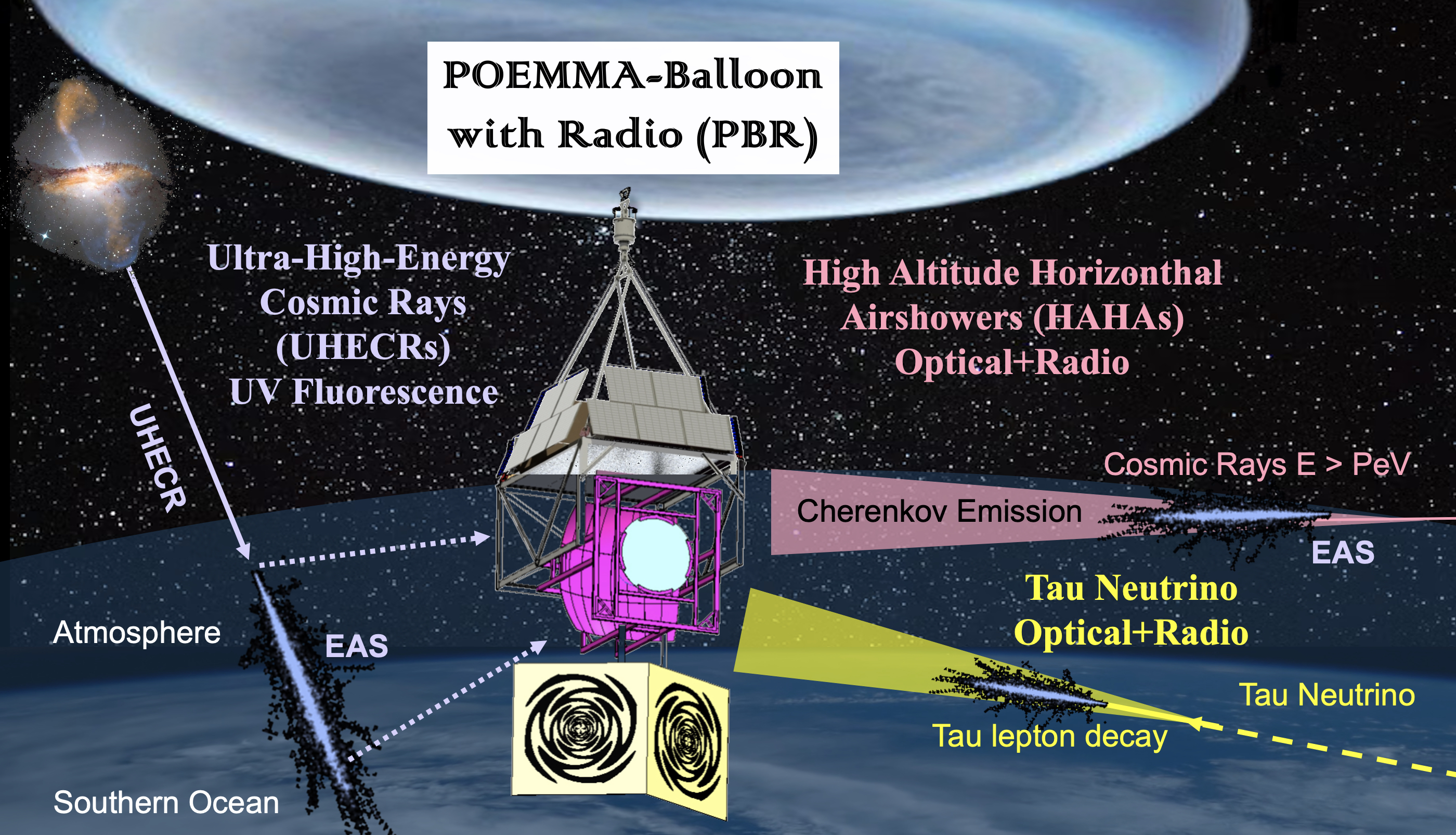}
    \caption{Main scientific goals of PBR: observation of UHECRs via fluorescence from above; study of HAHAs; and follow-up on astrophysical events in the search for astrophysical neutrinos.}
    \label{fig:ScienceSketch}
\end{figure}

Stratospheric balloon missions are an optimal intermediate step toward a space-based mission. They allow for improvements in technological readiness, validation of targeted detection techniques, and collection of initial scientific observations, all with reduced risk, time commitment, and cost compared to a fully realized space mission. The POEMMA-Balloon with Radio (PBR) is the most advanced stratospheric balloon experiment designed by the JEM-EUSO collaboration. It employs a hybrid focal surface in a wide-field Schmidt telescope, following the POEMMA design, and builds upon the successes of previous balloon missions, such as EUSO-SPB1 \cite{JEM-EUSO:2023ypf} and EUSO-SPB2 \cite{Adams:2025owi}. Despite challenges with balloon stability, these earlier missions provided valuable insights and technical advancements that have made PBR feasible. PBR will, for the first time, be equipped with a radio detector to provide a secondary, independent measurement channel for the phenomena under investigation. In general, the main scientific goals of PBR are threefold and illustrated in Fig.~\ref{fig:ScienceSketch}:

\begin{enumerate}
    \item \textbf{Investigating the origin of UHECRs:} PBR will observe, for the first time, the fluorescence emission of EASs produced by UHECRs from sub-orbital altitudes. This goal builds on the efforts of EUSO-SPB1 and EUSO-SPB2, which faced challenges in achieving stable, long-duration flights.
    \item \textbf{Studying High-Altitude Horizontal Air Showers (HAHAs):} PBR will observe a significant number of HAHAs, providing valuable data to study this rarely measured phenomenon. PBR will be able to use HAHAs to measure the cosmic-ray spectrum and discriminate cosmic-ray composition around PeV energies. In addition, PBR will detect the simultaneous optical Cherenkov and radio emissions of EASs, providing a multi-hybrid dataset crucial for understanding these high-energy events.
    \item \textbf{Searching for astrophysical neutrinos:} PBR will search for neutrinos from multi-messenger events, such as gamma-ray bursts and binary coalescences of compact objects. By detecting EASs arriving from below the Earth's limb, PBR aims to identify $\tau$-neutrinos interacting within the Earth that produce $\tau$-leptons, which decay and generate EASs.
\end{enumerate}

%% file: PBR-Payload.tex
\section{PBR-Payload and its expected performance}
\label{sec:payload}

\begin{figure}
    \centering
    \includegraphics[width=.75\linewidth]{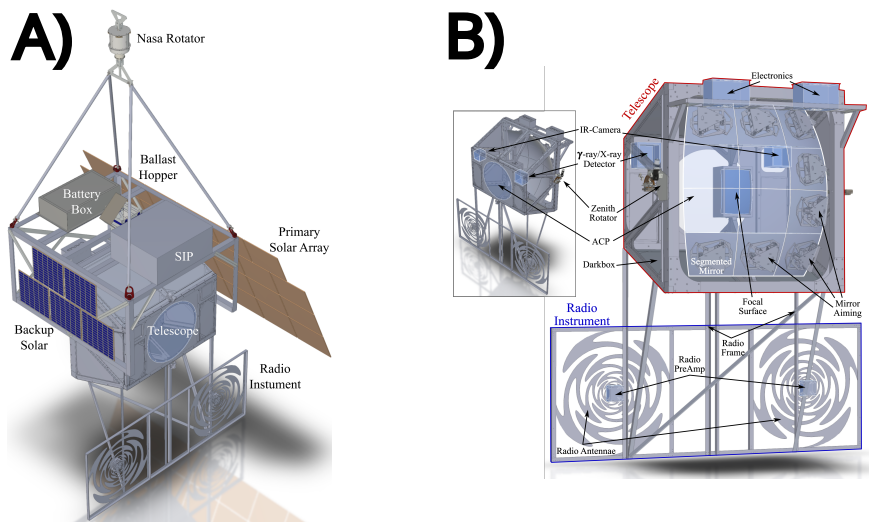}
    \caption{Left: Design of the full PBR payload, including SPB equipment. Right: Detailed design drawing of the two main PBR detectors: the Schmidt telescope with its hybrid focal surface and the radio antennas mounted beneath.}
    \label{fig:payload}
\end{figure}

The left panel of Fig.~\ref{fig:payload} shows the design of the PBR mission, including its configuration for flight as a NASA Super Pressure Balloon (SPB) from Wanaka, New Zealand, in the first half of 2027. The payload is expected to circulate over the Southern Ocean for more than 20 days, observing the atmosphere below and acquiring nighttime data from an altitude of approximately 33~km. The major components of PBR are a large, tiltable telescope housing a hybrid focal surface, and a Radio Instrument (RI) comprising two antennas mounted beneath the telescope.

A NASA rotator will allow the instrument to aim at night toward azimuthal directions of astrophysical interest, using feedback from a differential Global Positioning System (GPS). A custom tilt mechanism enables adjustment of the telescope’s elevation angle from nadir to $12^{\circ}$ above the horizontal. This range provides the lowest possible energy threshold (nadir) for UHECR observations, the capability to point near the limb for astrophysical event follow-up, and the ability to conduct periodic in-situ checks of optical focus across the camera’s FoV using stars.

The telescope itself (see right panel of Fig.~\ref{fig:payload}) is a modified Schmidt design with a \unit[1.1]{m} diameter aperture, covered by an aspheric corrector plate made of PMMA to correct for spherical aberration. Its primary mirror is composed of 12 individual segments with a radius of curvature of \unit[1.6]{m}. This optical system provides an expected point-spread function (95\% containment) of 3~mm in diameter, with a field of view of approximately $36^{\circ} \times 30^{\circ}$ at its focal surface. The hybrid camera located there consists of two parts: the fluorescence camera (FC), optimized for UHECR detection via the fluorescence technique, and the Cherenkov camera (CC), designed to measure the Cherenkov light from Earth-skimming neutrinos or high-altitude horizontal air showers. Both cameras and their expected performance will be discussed in the following sections, while additional details of the mechanical design are provided in \cite{Eric:ICRC}.

\subsection{Fluorescence Camera (FC)}

\begin{figure}[!ht]
    \centering
        \includegraphics[width=1.\linewidth]{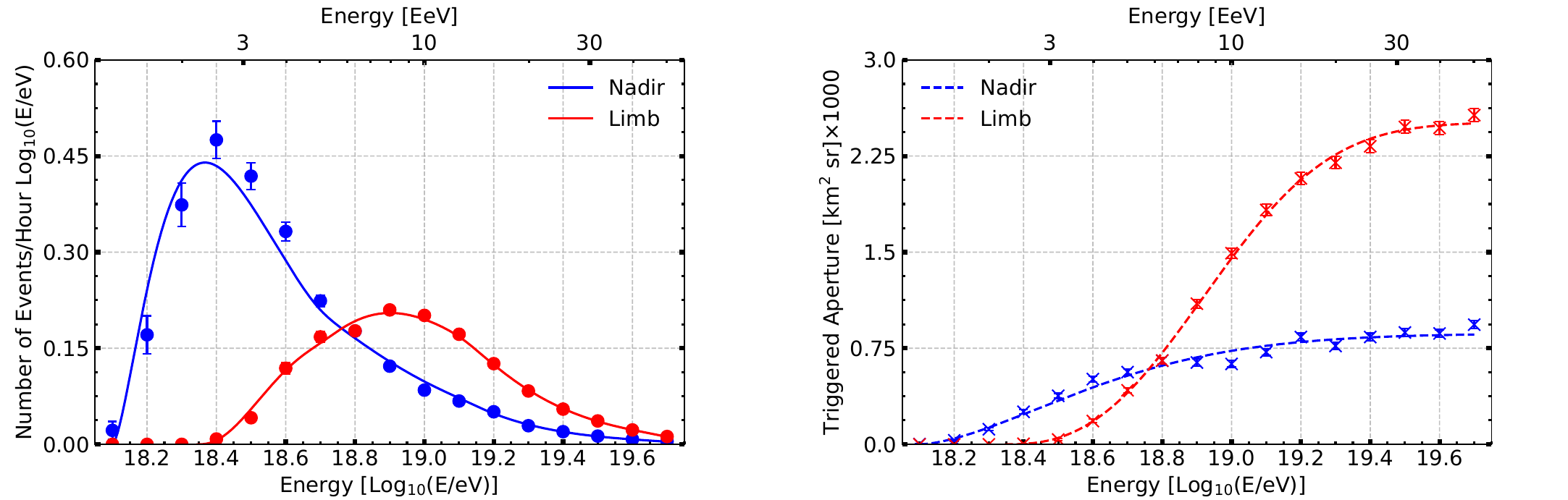}
    \caption{Left: Preliminary event rate of PBR for two pointing scenarios — nadir (blue) and limb (red). Right: Triggered aperture of PBR for both pointing scenarios.}
    \label{fig:FCperformance}
\end{figure}

With an integration time of \unit[1]{$\mu$s} and a double-pulse resolution of approximately \unit[10]{ns} in single photoelectron counting mode, the FC is ideally suited to measure extensive air showers (EASs) initiated by UHECRs. To reduce background light, a BG3 filter constrains the wavelength range between 290 and \unit[430]{nm}. Mechanically and electronically, the FC follows the same modular approach as previous missions: four 64-channel multi-anode photomultiplier tubes are combined into one elementary cell, and nine such cells form one Photo Detection Module (PDM) with 2304 individual pixels. PBR's FC will have four PDMs, providing a total FoV of approximately $24^{\circ} \times 24^{\circ}$, with an instantaneous FoV of $0.2^{\circ}$ per pixel. Based on this design and the experience of previous missions, a large-scale simulation study using the EUSO-\mbox{$\overline{\rm Off}$\hspace{.05em}\raisebox{.3ex}{$\underline{\rm line}\enspace$}}framework \cite{JEM-EUSO:2023fyg} has been conducted, revealing the expected FC performance (Fig.~\ref{fig:FCperformance}). Details of the simulation approach, developed for the EUSO-SPB1 and EUSO-SPB2 missions, can be found in \cite{Adams:2024gsj}.

Two pointing configurations were considered. In the nadir case, the lowest energy threshold of $\sim$\unit[1.8]{EeV} results in a rate of roughly 0.2 events per observation hour. In the limb case, where the top of the FC FoV is aligned with the Earth's limb, the geometric aperture is significantly increased but the energy threshold rises to \unit[4]{EeV}, leading to an event rate of $\sim$0.07 events per observation hour. Simulations indicate that over 10\% of events will be of high quality — defined as events that allow reconstruction of the arrival direction, energy, and potentially the composition of the primary cosmic ray. An in-depth discussion is provided in \cite{Francesco:ICRC}. This first EAS measurement from above via the fluorescence method will raise the technology readiness level (TRL) of this technique to~6, a requirement for future space missions.

\subsection{Cherenkov Camera (CC)}

The 2048-pixel CC has a modular design comprising four PDMs, each consisting of eight SiPM arrays (Hamamatsu S13361-3050). Each array contains 64 channels with an active area of $3\times\unit[3]{mm^2}$ per pixel, and is sensitive to wavelengths from 320 to \unit[900]{nm}. To optimize optical performance, the arrays are mounted on a curved structure approximating the spherical curvature of the telescope’s focal surface. The total FoV spans $12^{\circ} \times 6^{\circ}$, with each pixel covering $0.2^{\circ}$. To reduce noise triggers, a bi-focalizing optical system — requiring local and temporal coincidence — is being developed for installation in front of the camera.

Two digitizing electronics options are currently being evaluated in parallel: RadioRoc and the MIZAR ASIC (developed by the INFN group in Turin, Italy), the latter offering a sampling frequency of \unit[200]{MHz}. Further discussion is provided in \cite{Valentina:ICRC}.

\begin{figure}[!ht]
    \centering
        \includegraphics[width=1.\linewidth]{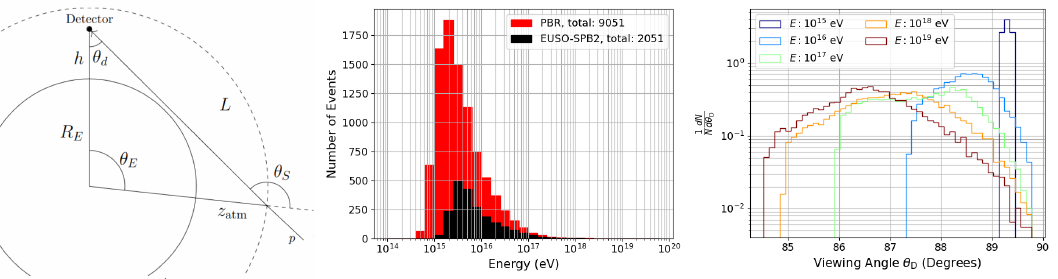}
    \caption{Left: Example HAHA trajectory illustrating the large atmospheric distances traversed. Center: Expected event rate for a 30-day flight (red: PBR, black: EUSO-SPB2). Right: Angular distribution for different primary cosmic-ray energies.}
    \label{fig:HAHA}
\end{figure}

This CC design enables PBR to study High-Altitude Horizontal Air Showers (HAHAs) when pointed above the limb and to search for astrophysical neutrinos when pointed below the limb. HAHAs not only serve as a guaranteed proxy signal for evaluating and refining the detection technique but also offer unique science potential, as they develop in rarefied atmosphere above \unit[20]{km}. The left panel of Fig.~\ref{fig:HAHA} shows sample trajectories.

The center panel of Fig.~\ref{fig:HAHA} shows an estimated HAHA event rate of $\sim$60 events per hour with an energy threshold of \unit[500]{TeV}, obtained using EASCherSim\footnote{https://gitlab.com/c4341/easchersim} \cite{Cummings:2020ycz, Cummings:2021bhg} for a 20\% duty cycle over a 30-day flight. A geometric energy filter can be applied — exploiting the fact that only high-energy events can span a wide angular range, while lower-energy (and more frequent) events are concentrated in the upper CC field — to estimate event energies. For the first time, these optical measurements will be combined with coincident radio measurements from the instrument described in Sec.~\ref{subsec:RI}, potentially providing an additional handle on primary cosmic-ray composition.

When pointed below the limb, the CC and RI are sensitive to Cherenkov radiation from upward-going EASs initiated by Earth-skimming neutrinos (primarily $\nu_{\tau}$), making PBR a pioneering instrument for studying astrophysical neutrino sources. Such detections would yield insights into cosmic-ray acceleration and interaction environments. Together with photons and gravitational waves, VHENs will help to characterize UHE accelerators (see \cite{Guepin:2022qpl}).

Due to the limited mission duration (even assuming 100 days) and the narrow FoV, PBR is expected to have limited sensitivity to the diffuse cosmogenic neutrino flux, even under optimistic UHECR source-evolution scenarios, consistent with current IceCube and Auger results \cite{Cummings:2020ycz}.

PBR’s greatest strength is its ability to repoint toward astrophysical events after receiving alerts, enabling time-sensitive observations of transients such as supernovae, binary neutron star mergers, tidal disruption events, blazar flares, and gamma-ray bursts.

\begin{figure}[!ht]
    \centering
    \raisebox{-.5\height}{\subfloat{
        \includegraphics[width=0.4\linewidth]{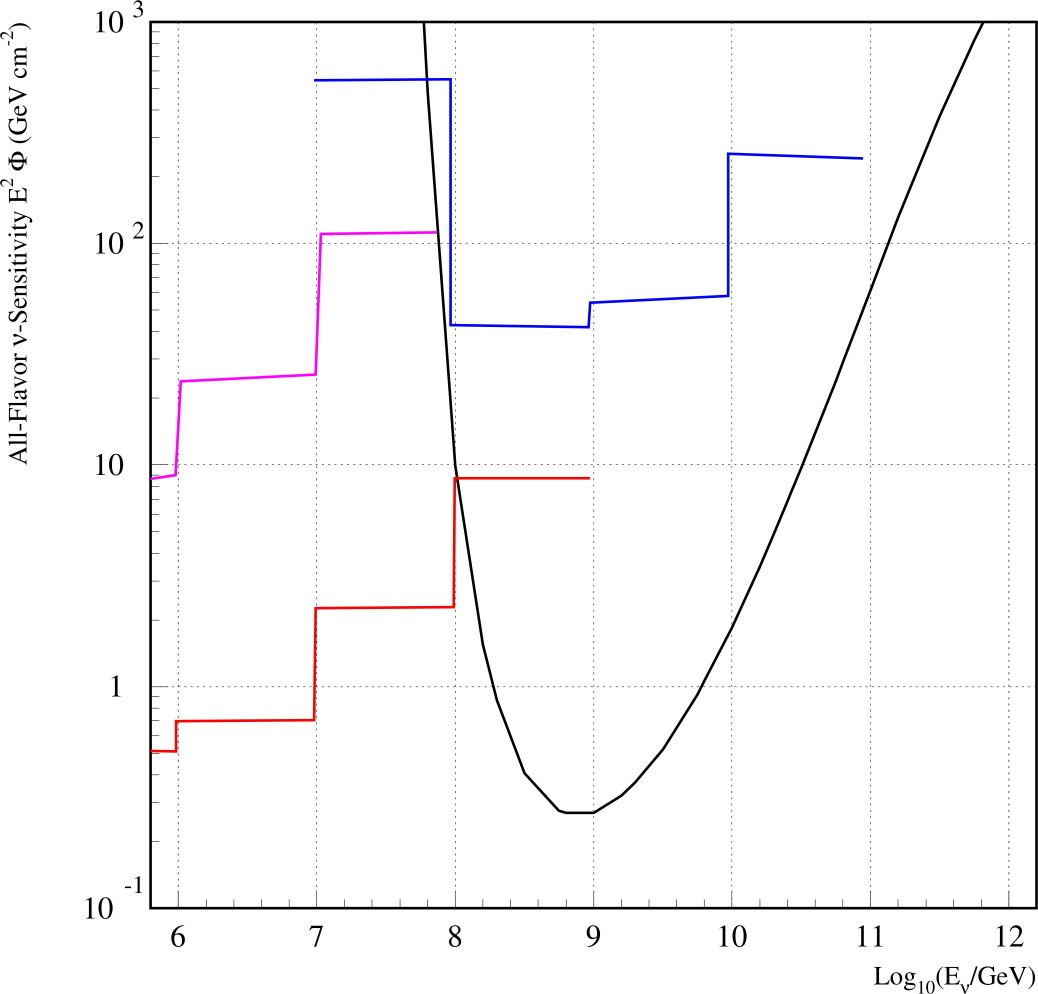}
        \label{fig:sub1}
    }}
    \hfill
    \raisebox{-.4\height}{\subfloat{
        \includegraphics[width=0.48\linewidth,trim = 6.0cm 6.0cm 6.0cm 6.25cm,clip]{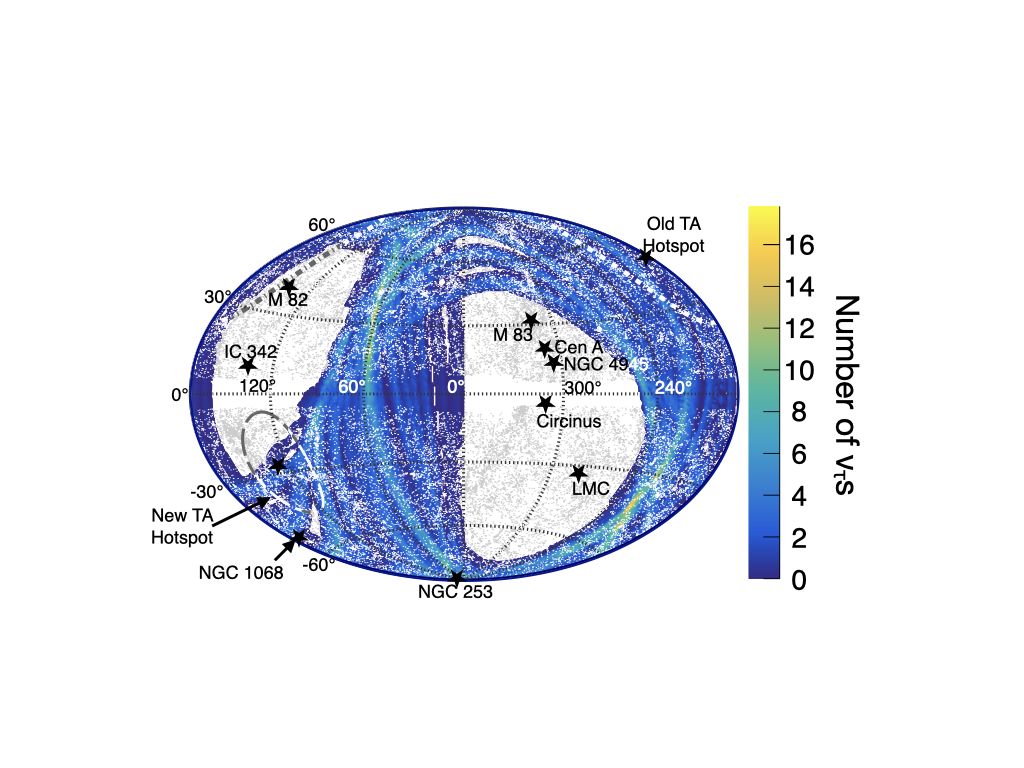}
        \label{fig:sub2}
    }}
    \caption{Left: Optimal 1000-s sensitivity for an April 2023 flight (black), compared with Auger (blue), IceCube (red), and ANTARES (magenta) limits for GW170817. Right: PBR's 100-day average number of $\nu_{\tau}$ events as a function of source location, assuming a BNS merger model~\cite{Fang:2017tla} at \unit[3]{Mpc} distance. Black stars indicate nearby sources.}
    \label{fig:ToO}
\end{figure}

The left panel of Fig.~\ref{fig:ToO} shows the strength of Target-of-Opportunity (ToO) follow-up with PBR, presenting all-flavor 90\% CL sensitivities for optimal 1000-s ToO observations using the $\nu$-SpaceSim software \cite{John:ICRC}. For comparison, limits from IceCube, Auger, and ANTARES for GW170817 are also shown.

The right panel presents a sky map color-coded by time-averaged effective area, assuming a 100-day April 2027 mission, accounting for Sun and Moon constraints. Full methodological details can be found in \cite{Venters:2019xwi}.

\subsection{Radio Instrument}
\label{subsec:RI}

PBR's Radio Instrument (RI) consists of two dual-polarized sinuous antennas, based on the design of PUEO's Low Frequency (LF) instrument, and optimized for air-shower detection \cite{Abarr_2021}. It provides an independent measurement channel to the Cherenkov Camera (CC) with identical scientific objectives. The antennas are scaled to diameters of 60$^{\prime\prime}$, offering a broadband 5~dBi response from 50--550~MHz. They are mounted 1~m apart, facing the same direction, to achieve a beamformed gain of approximately 6.5~dBi, enable azimuthal reconstruction of signals, and provide a field of view of $\pm30^{\circ}$.

The antennas are fabricated from copper to improve gain and are printed on an FR4 substrate, which is mounted to a lightweight aramid honeycomb structure with an FR4 backplate for mechanical rigidity and to support the mounting of electronics. The RI is positioned beneath PBR's optical cameras and aligned with the optical axis, ensuring overlapping fields of view following zenith and azimuth rotations of the telescope.

The RI primarily operates using an external trigger initiated by CC detections, enabling hybrid observations of air showers, but it also includes a dedicated radio‑only self‑trigger—intended mainly for daytime operation when the CC is offline. Design and development of the optimized RI electronics are currently underway, building on in-depth testing of PUEO's LF system, with the first hardware tests expected within the coming months. Simulations to evaluate the RI's standalone sensitivity to atmosphere-skimming air showers are also in progress.

%% file: Summary.tex
\section{Summary}
\label{sec:summary}

The next-generation stratospheric balloon payload currently under construction by the JEM-EUSO collaboration is PBR. This mission is a crucial step toward making space-based observations of UHECRs and VHENs a reality in the near future. For the first time, it will measure fluorescence light from UHECRs from above, with the additional aims of studying high-altitude horizontal air showers and detecting Earth-skimming neutrinos by following up on astrophysical event alerts. In addition, the implementation of the POEMMA-design hybrid focal surface will enhance its technical readiness for future space missions.

The addition of radio antennas will allow PBR, for the first time, to combine measurements of the radio and optical Cherenkov signals, providing an independent measurement channel.

The target launch for PBR is scheduled for spring 2027 from Wanaka, New Zealand, as a NASA Super Pressure Balloon payload, with a planned mission duration exceeding 20 days.\\

\footnotesize{\textit{Acknowledgment:}
The authors would like to acknowledge the support by NASA award 80NSSC22K1488 and 80NSSC24K1780, by the French space agency CNES and the Italian Space agency ASI. The work is supported by OP JAC financed by ESIF and the MEYS CZ.02.01.01/00/22\_008/0004596. We gratefully acknowledge the collaboration and expert advice provided by the PUEO collaboration. This research used resources of the National Energy Research Scientific Computing Center (NERSC), a U.S. Department of Energy Office of Science User Facility operated under Contract No. DE-AC02-05CH11231. We acknowledge the NASA Balloon Program Office and the Columbia Scientific Balloon Facility and staff for support. We also acknowledge the invaluable contributions of the administrative and technical staffs at our home institutions.
}

%% file: JEM-EUSO_Authors_August2025.tex
    \newpage
{\Large\bf Full Authors list: The JEM-EUSO Collaboration}

\begin{sloppypar}
{\small \noindent
M.~Abdullahi$^{ep,er}$              
M.~Abrate$^{ek,el}$,                
J.H.~Adams Jr.$^{ld}$,              
D.~Allard$^{cb}$,                   
P.~Alldredge$^{ld}$,                
R.~Aloisio$^{ep,er}$,               
R.~Ammendola$^{ei}$,                
A.~Anastasio$^{ef}$,                
L.~Anchordoqui$^{le}$,              
V.~Andreoli$^{ek,el}$
% ...
}
\end{sloppypar}
\vspace*{.3cm}

{ \footnotesize
\noindent
%
% ...
$^{cb}$ Universit\'e de Paris, AstroParticule et Cosmologie (CNRS), Paris, France\\
% ...
$^{ef}$ Istituto Nazionale di Fisica Nucleare (INFN), Sezione di Napoli, Naples, Italy\\
% ...
$^{ei}$ Istituto Nazionale di Fisica Nucleare (INFN), Sezione di Roma Tor Vergata, Rome, Italy\\
% ...
$^{ek}$ Istituto Nazionale di Fisica Nucleare (INFN), Sezione di Torino, Turin, Italy\\
$^{el}$ Universit\`a di Torino, Dipartimento di Fisica, Turin, Italy\\
% ...
$^{ep}$ Gran Sasso Science Institute (GSSI), L’Aquila, Italy\\
$^{er}$ Istituto Nazionale di Fisica Nucleare (INFN), Laboratori Nazionali del Gran Sasso, Assergi, Italy\\
% ...
$^{ld}$ University of Alabama in Huntsville, Huntsville, AL, USA\\
$^{le}$ City University of New York (CUNY), Lehman College, Bronx, NY, USA\\
% ...
}

%% file: ICRC2025_PBRoverview.bbl
\providecommand{\href}[2]{#2}\begingroup\raggedright\begin{thebibliography}{10}

\bibitem{Coleman:2022abf}
A.~Coleman et~al., \emph{{Ultra high energy cosmic rays The intersection of the Cosmic and Energy Frontiers}}, \href{https://doi.org/10.1016/j.astropartphys.2023.102819}{\emph{Astropart. Phys.} {\bfseries 149} (2023) 102819} [\href{https://arxiv.org/abs/2205.05845}{{\ttfamily 2205.05845}}].

\bibitem{POEMMA:2020ykm}
{\scshape POEMMA} collaboration, \emph{{The POEMMA (Probe of Extreme Multi-Messenger Astrophysics) observatory}}, \href{https://doi.org/10.1088/1475-7516/2021/06/007}{\emph{JCAP} {\bfseries 06} (2021) 007} [\href{https://arxiv.org/abs/2012.07945}{{\ttfamily 2012.07945}}].

\bibitem{Zbigniew:ICRC}
{\scshape JEM-EUSO} collaboration, \emph{{From Ground to Space: An Overview of the JEM-EUSO Program for the Study of UHECRs and Astrophysical Neutrinos}}, {\emph{PoS} {\bfseries ICRC2025} (2025) 360}.

\bibitem{JEM-EUSO:2023ypf}
{\scshape JEM-EUSO} collaboration, \emph{{EUSO-SPB1 mission and science}}, \href{https://doi.org/10.1016/j.astropartphys.2023.102891}{\emph{Astropart. Phys.} {\bfseries 154} (2024) 102891} [\href{https://arxiv.org/abs/2401.06525}{{\ttfamily 2401.06525}}].

\bibitem{Adams:2025owi}
{\scshape JEM-EUSO} collaboration, \emph{{The Extreme Universe Observatory on a Super-Pressure Balloon II: Mission, Payload, and Flight}},  \href{https://arxiv.org/abs/2505.20762}{{\ttfamily 2505.20762}}.

\bibitem{Eric:ICRC}
{\scshape JEM-EUSO} collaboration, \emph{{The Optical and Mechanical Design of POEMMA Balloon with Radio}}, {\emph{PoS} {\bfseries ICRC2025} (2025) 332}.

\bibitem{JEM-EUSO:2023fyg}
{\scshape JEM-EUSO} collaboration, \emph{{EUSO-Offline: A comprehensive simulation and analysis framework}}, \href{https://doi.org/10.1088/1748-0221/19/01/P01007}{\emph{JINST} {\bfseries 19} (2024) P01007} [\href{https://arxiv.org/abs/2309.02577}{{\ttfamily 2309.02577}}].

\bibitem{Adams:2024gsj}
{\scshape JEM-EUSO} collaboration, \emph{{The EUSO-SPB2 fluorescence telescope for the detection of Ultra-High Energy Cosmic Rays}}, \href{https://doi.org/10.1016/j.astropartphys.2024.103046}{\emph{Astropart. Phys.} {\bfseries 165} (2025) 103046} [\href{https://arxiv.org/abs/2406.13673}{{\ttfamily 2406.13673}}].

\bibitem{Francesco:ICRC}
{\scshape JEM-EUSO} collaboration, \emph{{The Flourescence Camera for the PBR mission}}, {\emph{PoS} {\bfseries ICRC2025} (2025) 209}.

\bibitem{Valentina:ICRC}
{\scshape JEM-EUSO} collaboration, \emph{{The Cherenkov Camera for the PBR mission}}, {\emph{PoS} {\bfseries ICRC2025} (2025) }.

\bibitem{Cummings:2020ycz}
A.L.~Cummings, R.~Aloisio and J.F.~Krizmanic, \emph{{Modeling of the Tau and Muon Neutrino-induced Optical Cherenkov Signals from Upward-moving Extensive Air Showers}}, \href{https://doi.org/10.1103/PhysRevD.103.043017}{\emph{Phys. Rev. D} {\bfseries 103} (2021) 043017} [\href{https://arxiv.org/abs/2011.09869}{{\ttfamily 2011.09869}}].

\bibitem{Cummings:2021bhg}
A.~Cummings, R.~Aloisio, J.~Eser and J.~Krizmanic, \emph{{Modeling the optical Cherenkov signals by cosmic ray extensive air showers directly observed from suborbital and orbital altitudes}}, \href{https://doi.org/10.1103/PhysRevD.104.063029}{\emph{Phys. Rev. D} {\bfseries 104} (2021) 063029} [\href{https://arxiv.org/abs/2105.03255}{{\ttfamily 2105.03255}}].

\bibitem{Guepin:2022qpl}
C.~Gu\'epin, K.~Kotera and F.~Oikonomou, \emph{{High-energy neutrino transients and the future of multi-messenger astronomy}}, \href{https://doi.org/10.1038/s42254-022-00504-9}{\emph{Nature Rev. Phys.} {\bfseries 4} (2022) 697} [\href{https://arxiv.org/abs/2207.12205}{{\ttfamily 2207.12205}}].

\bibitem{Fang:2017tla}
K.~Fang and B.D.~Metzger, \emph{{High-Energy Neutrinos from Millisecond Magnetars formed from the Merger of Binary Neutron Stars}}, \href{https://doi.org/10.3847/1538-4357/aa8b6a}{\emph{Astrophys. J.} {\bfseries 849} (2017) 153} [\href{https://arxiv.org/abs/1707.04263}{{\ttfamily 1707.04263}}].

\bibitem{John:ICRC}
{\scshape $\nu$SpaceSim} collaboration, \emph{{$\nu$SpaceSim: A Comprehensive Simulation Package for Modeling the Measurement of Cosmic Neutrinos using the Earth as the Neutrino Target and Space-based Detectors}}, {\emph{PoS} {\bfseries ICRC2025} (2025) 1082}.

\bibitem{Venters:2019xwi}
T.M.~Venters, M.H.~Reno, J.F.~Krizmanic, L.A.~Anchordoqui, C.~Gu\'epin and A.V.~Olinto, \emph{{POEMMA's Target of Opportunity Sensitivity to Cosmic Neutrino Transient Sources}}, \href{https://doi.org/10.1103/PhysRevD.102.123013}{\emph{Phys. Rev. D} {\bfseries 102} (2020) 123013} [\href{https://arxiv.org/abs/1906.07209}{{\ttfamily 1906.07209}}].

\bibitem{Abarr_2021}
Q.~Abarr, P.~Allison, J.~Ammerman~Yebra, J.~Alvarez-Muñiz, J.~Beatty, D.~Besson et~al., \emph{{The Payload for Ultrahigh Energy Observations (PUEO): a white paper}}, \href{https://doi.org/10.1088/1748-0221/16/08/p08035}{\emph{Journal of Instrumentation} {\bfseries 16} (2021) P08035}.

\end{thebibliography}\endgroup
